# Cassini/VIMS hyperspectral observations of the HUYGENS landing site on Titan

S. Rodriguez[1*], S. Le Mouélic[1], C. Sotin[1], H. Clénet[1], R.N. Clark[2], B. Buratti[3], R.H. Brown[4], T.B. McCord[5], P.D. Nicholson[6], K.H. Baines[3] and the VIMS science team

[1] *Laboratoire de Planétologie et Géodynamique, UMR CNRS 6112, Université de Nantes, France.*
[2] *USGS, Denver Federal Center, Denver CO, USA*
[3] *JPL, California Institute of Technology, Pasadena, USA*
[4] *Lunar and Planetary Lab and Stewart Observatory, University of Arizona, Tucson, USA*
[5] *Department of Earth and Space Sciences, University of Washington, USA*
[6] *Cornell University, Astronomy Department, USA*

Number of MS Word pages: 38
Number of Table(s): 1
Number of figure(s): 11

Running Head: VIMS images of the Huygens probe landing site on Titan

[*] Correspond with:
Sébastien Rodriguez
CNRS/Université de Nantes
Laboratoire de Planetologie et Geodynamique
2 rue de la Houssiniere BP 92208
44322 Nantes cedex 3 France
ph. +33 (0)2 51 12 54 67
fax +33 (0)2 51 12 52 68
Sebastien.Rodriguez@univ-nantes.fr

**Abstract**

Titan is one of the primary scientific objectives of the NASA-ESA-ASI Cassini-Huygens mission. Scattering by haze particles in Titan's atmosphere and numerous methane absorptions dramatically veil Titan's surface in the visible range, though it can be studied more easily in some narrow infrared windows. The VIMS instrument (Visual and Infrared Mapping Spectrometer) onboard the Cassini spacecraft successfully imaged its surface in the atmospheric windows, taking hyperspectral images in the range 0.4 to 5.2 µm. On 26 October (TA flyby) and 13 December 2004 (TB flyby), the Cassini-Huygens mission flew over Titan at an altitude lower than 1200 km at closest approach. We report here on the analysis of VIMS images of the Huygens landing site acquired at TA and TB, with a spatial resolution ranging from 16 to14.4 kilometers per pixel. The pure atmospheric backscattering component is corrected by using both an empirical method and a first order theoretical model. Both approaches provide consistent results. After the removal of scattering, ratio images reveal subtle surface heterogeneities. A particularly contrasted structure appears in ratio images involving the 1.59 µm and 2.03 µm images north of the Huygens landing site. Although pure water ice cannot be the only component exposed at Titan's surface, this area is consistent with a local enrichment in exposed water ice and seems to be consistent with DISR/Huygens images and spectra interpretations. The images show also a morphological structure that can be interpreted as a 150 km diameter impact crater with a central peak.

# 1. Introduction.

Titan is the second largest satellite in the Solar System, after Jupiter's moon Ganymede. This planet-like moon is the only one to possess a dense and extended atmosphere, which is primarily composed of $N_2$ with less than 5% $CH_4$ and 1% $H_2$. Titan's thick atmosphere is subject to an active chemistry induced by solar radiation, solar wind and charged particles from the Saturnian magnetosphere (**Yung *et al.*, 1984; Toublanc *et al.*, 1995; Lara *et al.*, 1996, 1998**). This chemistry results in the production of a great diversity of hydrocarbons and nitriles in the high stratosphere with increasing complexity and molecular weight. Settling in the atmosphere, chemical end products form an extensive system of organic aerosol haze responsible for Titan's characteristic orange colour and for hiding its surface at ultraviolet and visible wavelengths. Its surface is observable in some infrared atmospheric windows and at greater wavelengths in the case of an unclouded low atmosphere (**Smith *et al.*, 1996; Rodriguez *et al*., 2003**). Radar observations, microwave radiometry, and near-infrared studies revealed the presence of an heterogeneous surface plausibly composed of water ice and organic compounds in solid or liquid form (**Muhleman *et al.*, 1990; Coustenis *et al.*, 1995, 1997, 2001**), but the precise nature of Titan's surface remains unknown.

The joint NASA-ESA-ASI Cassini-Huygens mission reached the Saturnian system on July 1st 2004. Since this date, the Cassini spacecraft has been in orbit around Saturn, making an extensive survey of the ringed planet and its moons. With 44 planned flybys over the four years of the nominal mission, Titan is a primary target of the Cassini orbiter.

Cassini was accompanied by the European Huygens probe, entirely dedicated to Titan study. Huygens was released from its mothership Cassini on 25[th] December 2004 and successfully parachuted on Titan's surface on 14<sup>th</sup> January 2005, being the first probe to land in the outer Solar System. The DISR (Descent Imager/Spectral Radiometer) instrument onboard Huygens took a series of spectra and images of the ever-approaching surface as it dropped through Titan's atmosphere (**Tomasko *et al.*, 2002; 2005**). The high resolution images acquired before the landing (**Figure 1-a**) revealed high contrasted terrains probably related to local topography, with bright regions interpreted to be a few hundred meters higher than darker ones. Bright regions are covered by dark streaks likely to be drainage channels cut by rivers of methane leading through deltas to a dark lower lakebed. This interpretation is consistent with pebble-sized "rocks" seen on the DISR image of Titan's surface acquired as Huygens has landed. The shape of these rocks strongly suggests erosion by a fluid, indicating possible fluvial activity. As the probe did not land on liquids, it is probable that the DISR images only witness ancient liquid traces that are now dried out (**Tomasko *et al.*, 2005**). Huygens observations provide important ground truth that will greatly help to constrain observations from orbit.

The VIMS (Visual and Infrared Mapping Spectrometer) instrument onboard Cassini acquires hyperspectral images with two separate detectors, one in the visible ranging from 0.35 to 1.05 μm, and one in the infrared from 0.88 to 5.1 μm, with a total of 352 spectral channels (**Brown *et al.*, 2003**). VIMS has proved to be able to see through the haze and successfully take medium to high resolution images of Titan's surface in seven infrared spectral windows, where atmospheric methane absorptions are the weakest (**Sotin *et al.*, 2005**). We discuss here the first images of the Huygens landing site acquired by VIMS during the first two flybys of Titan that occurred in October and December 2004, before the Huygens probe successfully landed on Titan. We show that these images complement the high resolution images acquired by the DISR instrument and we discuss the morphological and spectral implications on the possible nature of Titan's surface around the Huygens landing site area.

**[Figure 1]**

## 2. Description of the sequences of observations.

### *2.1. VIMS images of the Huygens landing site.*

The first close flyby of Titan by the Cassini spacecraft (tagged TA) occurred on 26 October 2004. During this flyby, two 64x64 pixels images of the Huygens entry site were acquired by VIMS with time exposures of respectively 200 and 40 ms (cubes 1477490933 and 1477491859). **Figures 2b** and **2c** show the two raw images in the 2.03 µm band, which is one of the seven infrared spectral windows. The cubes were radiometrically calibrated through the standard VIMS calibration pipeline. **Figures 2e** and **2f** show the corresponding cylindrical reprojections on a Titan geographic map (north is up). The geometrical reconstruction of VIMS images is made possible by the knowledge of the spacecraft attitude and position with time. The long time exposure image exhibits significant distortions (**Figure 2b**), compared with the short time exposure image (**Figure 2c**). These distortions are due to the spacecraft motion during the acquisition, as Cassini turned from inertia wheels to thrusters, and have been corrected on the reprojected image (**Figure 2e**). Mean incidence, emergence and phase angles during these observations are respectively 35°, 30° and 13°. The long and short time exposure images were acquired when Cassini was 35000 km and 28000 km over Titan's surface, a few hours before TA closest approach. The spatial resolution ranges for the two images from 16.2 km to 14.4 km per pixel. At these resolutions, no significant changes are observed between the two acquisitions, which were made 15 minutes apart, suggesting that all features seen on these images belong to the surface.

**[Figure 2]**

The second flyby (tagged TB) occurred on 13 December 2004. Riding along with the ISS instrument, VIMS acquired ten 24x12 pixels cubes with a 160 ms time exposure in order to produce a 5x2 mosaic of the Huygens site (cubes 1481623711, 1481623853, 1481624032, 1481624171, 1481624349, 1481624482, 1481624672, 1481624812, 1481624991, and 1481625137, calibrated through the standard VIMS calibration pipeline – see **Figure 2d**). The resulting spatial resolution ranges from 19 to 14.7 km per pixel. During the acquisition the mean incidence (and emergence) angle varied from 45° (44°) for the lower left of the mosaic to 30° (24.5°) for the upper right. The phase angle is relatively constant over the mosaic, increasing slightly from 18° to 19°. Observation conditions were very close to the one at TA. On the **Figure 2g** is displayed the reprojected mosaic in the 2.03 µm VIMS band. **Figure 3** shows contours of the main I/F units of the TB 2.03µm image (**Figure 2g**) overlaying the TA 2.03 µm image (**Figures 2f**). The contours of the TB image match closely the I/F variations observed at TA, showing that at this spatial resolution (~15 km/pixel) no major change occurred during the 48 days separating the two encounters. This confirms that the main structures observed in this region are due to the surface only and not to transient atmospheric phenomena.

**[Figure 3]**

*2.2. Spectral characteristics of VIMS "Huygens" scenes.*

The VIMS instrument acquires hyperspectral images in two separate channels covering the ranges 0.3–1.05 µm (Visual channel) and 0.8–5.1 µm (IR channel) with nominal spectral sampling of 7.3 nm (96 bands) and 16.6 nm (256 bands), respectively. VIMS cubes of the Huygens landing site do not pre-sent major spectral differences. Scattering of solar radiation from particles in the atmosphere and strong absorptions by methane globally dominate the shape of VIMS Titan spectra, lowering the overall albedo level toward zero except in seven atmospheric windows (0.93, 1.08, 1.27, 1.59, 2.03, 2.75, and 5.0 μm).

[Figures 4a-b-c]

**Figure 4a** (**Figure 4b**: zoom between 0.8 and 2.2 μm) shows the most different spectra extracted from the long time exposure image cube acquired during TA. The two spectra are associated with the most contrasted types of terrains that can be found in the image: the spectrum with high I/F level corresponds to a pixel in the western bright patch and the spectrum with low I/F level to a pixel in the darkest north-east area. As expected, the two spectra are very similar in shape and the only differences are found in the atmospheric windows. At first order, the only spectral difference is the general level in the windows, where the observed I/F is thought to result predominantly from surface reflectance properties and topography. However, scattering by aerosols and weaker but not negligible methane absorption make the extraction of the surface contribution in these near-infrared windows challenging. It can theoretically be done by a complex and extensive modelling of haze scattering and methane absorption (**Griffith *et al.*, 1991, 2003; Coustenis *et al.*, 1995, 1997, 2001, 2005**).

In order to enhance the subtle differences between the "bright" and "dark" spectra presented in **Figure 4a**, we calculate the ratio "bright"/"dark" (**Figure 4c**). With the simple assumption that the absorption by Titan's atmosphere does not vary over the "Huygens" region, the spectral ratios should only be sensitive to surface reflectance, topographical variations, and scattering by aerosols.

All the windows display positive ratios with, in general, a smooth symmetrical shape with respect to the window centers. They do not exhibit fine structures that could have witnessed evidence of surface composition variations between bright and dark regions. Only the ratio in the 1.59 μm window still presents two small shoulders on both sides of the central peak at ~1.55 and ~1.62 μm and indicates possible surface absorption features present on bright area and less on dark. Pure $CO_2$ ice or $CO_2$ diluted in less than 5% of $H_2O$ displays a triplet of small bands in this spectral region at 1.54, 1.58, and 1.61 μm (**Quirico and Schmitt, 1997**) and could provide a good candidate. Differences in optical path length, or contrasts in topography between bright and dark terrains suggested by DISR/Huygens images (**Tomasko *et al.*, 2005**), could also explain the small persistent shoulders observed in the spectral ratio at 1.55 and 1.62 μm. Photons reflected on the highlands regions must have a shorter atmospheric path length than those reflected on lower regions, inducing a lower absorption by the atmospheric methane which is the major absorbent in the last hundreds of meters above the surface. DISR measurements near Titan's surface showed during Huygens probe descent great changes in the shape of spectra separated by only a few tens of meters in altitude (**Tomasko *et al.*, 2005**). Thus variations in methane absorption due to hundred meters of topography may be sufficient to be detected in VIMS spectra and may contribute (with surface composition) to the I/F level in Titan's spectral windows. We can say at this stage that the positive values of the spectral ratio that occur in all spectral windows (**Figure 4c**) is consistent with what Huygens DISR images suggest: bright regions are likely to be globally higher than dark regions.

As the individual spectra are dominated by the atmospheric contributions (absorption and scattering), the resulting ratio is globally close to unity, except in the windows at 0.93, 1.08, 1.27, 1.59, 2.03, and 2.75 μm and everywhere beyond 4 μm where the VIMS spectra have extremely a low I/F ratio and become very noisy. With a ratio value of 2, the 2.03 μm band is the most contrasted of all the atmospheric windows. Other windows below 2.03 μm have a ratio value that decreases with decreasing wavelength to an asymptotic value of one. This peculiar behaviour shows that VIMS spectral ratios are affected by backscattering by sub-micron atmospheric particles, with an efficiency that seems to decrease with increasing wavelength. It explains the decrease in contrast in VIMS images towards short wavelengths. This issue will be addressed in the next section.

## 3. First order removal of backscattering by aerosols.

### 3.1. Empirical correction of the additive backscattering component.

As seen in the previous section, the atmospheric aerosol layer affects VIMS observations of Titan's surface in the seven spectral windows. Scattering by aerosols adds to the pure surface signal a fraction of incident photons that interacted both with Titan's surface and atmosphere and a fraction that has never reached Titan's surface before being scattered to VIMS. This last scattering component acts on the data as a wavelength dependent offset in I/F values, totally independent of surface spectral properties.

The removal of the atmospheric contribution (aerosol scattering and gas absorptions) can be performed by using a high-fidelity radiative transfer code, but these codes require an *a priori* knowledge of the surface albedo (**Griffith *et al.*, 1991, 2003; Coustenis *et al.*, 1995, 1997, 2001, 2005**).

We propose here a first approach based on the removal of photons reflected by Titan's atmosphere without any interaction with the surface, already successfully tested to derive local topography in VIMS high resolution images (**Sotin *et al.*, 2005**). This highly simplified correction method treats only a fraction of the photons scattered by Titan's atmosphere, but it has the advantage that it requires no assumptions on absolute surface albedo.

We focus our correction on the seven wavelengths centred on the spectral windows, which are less affected by absorption than the other parts of Titan's spectrum. We deduced the reflective scattering contribution from aerosols for these seven wavelengths by using the following heuristic method: for each band $B_\lambda$, we compute the additive constant A which minimizes the standard deviation of the ratio $(B_\lambda+A)/(B_{5\mu m}+\varepsilon)$, where $B_{5\mu m}$ is the image at 5 µm. This method is similar to the one which was used by **Le Mouélic *et al.* (1999)** to derive the global offset of Clementine NIR infrared images at 1.1 µm of the Moon using the correlation with well calibrated Clementine UVVIS data at 1.0 µm. In the case of Titan, we assume that the aerosol scattering at 5µm is negligible, *i.e.*, $\varepsilon \ll 0$. This method is only valid when bands at $\lambda$ and 5 µm are strongly correlated. This is generally the case for Titan VIMS observations where we observe the same bright and dark areas in all spectral windows. Calculations were done for TA and TB cubes and the resulting values $A_\lambda$ for each spectral window are given in **Table 1** (columns marked "empirical method") and plotted in **Figure 5**.

*3.2. Comparison with a single scattering model and validation of the empirical method*.

The additive component due to scattering depends mainly on the haze properties, observing conditions and wavelength. Within the frame of a single scattering approximation, it is possible to quantify to first order the rise in albedo on VIMS images due to Titan's aerosol layer, thanks to the following equation giving the observed reflectance $R$ of a planetary body with an atmosphere (**Sobolev, 1975; Combes *et al.*, 1991**):

$$\frac{R}{\mu_0} = R_{surf}\left[1 - \tau_{atm}\left(\frac{1}{\mu} + \frac{1}{\mu_0}\right)\right] \qquad (1)$$

$$+ \omega_0 \frac{f(\gamma)}{4} \times \frac{\tau_{atm}}{\mu_0 \mu} \qquad (2)$$

$$+ \frac{R_{surf} \, \omega_0 \tau_{atm}}{2\mu_0} \int_0^1 p(\mu', \mu_0) d\mu' \qquad (3)$$

$$+ \frac{R_{surf} \, \omega_0 \tau_{atm}}{2\mu} \int_0^1 p(\mu, \mu_0') d\mu_0' \qquad (4)$$

$$+ R_{surf}^2 \, \omega_0 \tau_{atm} \int_0^1 d\mu' \int_0^1 p(-\mu', \mu_0') d\mu_0' \qquad (5)$$

where: **(1)** is the light reflected from the surface taking into account the extinction by atmospheric scatterers, **(2)** is the light scattered to the instrument before reaching the surface, **(3)** is the light reflected, then scattered by the atmosphere to the instrument, **(4)** is the light first scattered to the surface, then reflected to the instrument, and **(5)** is the light reflected, scattered by the atmosphere back to the surface, and reflected again in the direction of the instrument.

In the above equation, $R_{surf}$ is the albedo of the surface, $\tau_{atm}$ is the optical depth of atmospheric scatterers (dust or cloud particles) and $\omega_0$ is the single scattering albedo of the atmospheric scatterers. By definition, $\tau_{atm} = (\sigma_{abs} + \sigma_{sca})\rho H$ and $\omega_0 = \sigma_{sca}/(\sigma_{abs} + \sigma_{sca})$, $\sigma_{abs}$ and $\sigma_{sca}$ being the absorption and scattering cross-sections depending on the size and the optical constants of scatterers, $\rho$ the number density of scatterers, and $H$ the vertical extent of the scatterers layer. The observing geometry is introduced in the equation by the means of $\mu_0$, $\mu$ and $\alpha$, which correspond respectively to the cosine of the incidence angle, the cosine of the emergence angle and the phase angle. Finally, $f(\gamma)$ is the phase function, normalized to $4\pi$, where $\gamma = 180° - \alpha$ is the scattering angle in the scattering plane, and $p(\mu, \mu_0) = \frac{1}{2\pi} \int_0^{2\pi} f(\gamma) d\phi$ is the corresponding phase integral.

The additive contribution of atmospheric scattering in $R$ can be estimated from term **(2)** of the above equation which contains only atmospheric backscattering. Considering Cassini averaged attitude conditions during the Huygens landing site observations at TA ($\tilde{\mu}_0 = 0.82$, $\tilde{\mu} = 0.87$, $\tilde{\alpha} = 13°$) and TB ($\tilde{\mu}_0 = 0.79$, $\tilde{\mu} = 0.82$, $\tilde{\alpha} = 18°$), we computed values of **(2)** over the VIMS infrared wavelength range for a set of aerosol layer parameters which are characteristic of Titan, with the further simplifying assumption of only one homogeneous haze layer composed of single sized particles. $\tau_{atm}$ and $\omega_0$ are estimated using Rayleigh and Mie theories. We use the optical constants measured on laboratory analogs of Titan's aerosols by **Khare et al. (1984)**. The transition from Rayleigh to Mie regimes occurs when the ratio between the scatterers' radius and wavelength becomes greater than 0.05 (**Ishimaru, 1997**). In the Rayleigh regime, we use the analytical Rayleigh phase function, defined as follows: $f(\gamma) = \frac{3}{4}(1 + \cos^2(\gamma))$. When in the Mie regime, we use the Henyey-Greenstein phase function, allowing simple numerical calculations by the use of a single model parameter $g$, with $g$ known as the Henyey-Greenstein asymmetry factor, ranging from -1 to +1, with negative value representing dominantly backscattering and positive values forward scattering behaviour, respectively (**Combes et al., 1991**). The only parameters of the model are the aerosol size, number density and asymmetry factor, and the aerosol layer thickness.

The wavelength dependence of the empirical additive offset we previously derived from VIMS images of the Huygens landing site (see **section 3.1**) is well reproduced by the model (**Figure 5**). The best fit is found for aerosol parameters (size, number density, layer thickness and asymmetry factor) fully consistent with previous estimates derived from Voyager observations or pure microphysical calculations (**Rages & Pollack, 1983; McKay *et al.*, 1989; Cabane *et al.*, 1992**). These values are $r = 0.18$ µm, $\rho = 100 \, \text{cm}^{-3}$, $H = 200$ km, and $g = 0.4$ (particles scattering mostly in the forward direction). Although our values for aerosol radius and density number are derived from a very simple model, they are also consistent with those inferred from the DISR experiment onboard Huygens (**Tomasko *et al.*, 2005**). The same aerosol parameters were found for both TA and TB flybys. **Figure 5** presents the empirical offset curves retrieved for TA and TB Huygens images along with the best fits from the model calculations. Besides, our model allows us to compute the aerosol haze transmittance versus wavelength and to verify that the modelled layer is almost transparent at high wavelengths (see **Table 1**). All the values from our calculations are summarized in **Table 1** (columns marked "Modelling").

[Figure 5]

[Table 1]

**4. Analysis of TA and TB Huygens landing site observations using band ratios corrected from haze backscattering.**

*4.1. Method.*

Band ratios are particularly useful to cancel out the effects of albedo and illuminating conditions and complete the previous analysis initiated with single spectra by adding the spatial dimension. In the case of Titan, if bands are selected among the atmospheric windows, their ratios should empha-size subtle spectral variations of Titan's surface. In **Figure 6**, we present two reprojected images at 1.08 and 2.03 µm extracted from the long time exposure cube acquired above the Huygens landing site at TA and a set of 12 ratios, uncorrected for aerosol scattering, and selected among the bands centered on Titan's atmospheric windows.

The ratios by short wavelength images (1.08, 1.27, and to a lower degree 1.59 µm) still display the same structures as single band images (see for example in **Figure 6** the 2.03 µm reference image in comparison with the first two rows of ratio images: 1.27/1.08 µm, 1.59/1.08 µm, 2.03/1.08 µm, 2.79/1.08 µm, 1.59/1.27 µm, 2.03/1.27 µm, 2.79/1.27 µm, and 2.03/1.59 µm). This strong correlation to the albedo can be attributed to the presence of a strong additive scattering contribution from atmospheric suspended particles, which scatter solar light more efficiently at low near-infrared wavelengths. Titan haze particles blur surface images at low wavelengths (1.08, 1.27, and to a lesser extent 1.59 µm), as shown by the 1.08 µm image (**Figure 6**), and affect all the ratios using these bands.

New ratio images including the subtraction of aerosols backscattering contribution and correction for atmospheric extinction described in **section 3** are presented in **Figure 7**. As expected, the correlation with the albedo has been almost completely removed.

**[Figures 6 and 7]**

*4.2. Morphological analysis.*

The main I/F variations in the 1.08 and 2.03 µm images (**Figures 6 and 7** – top) are dominated by two contrasted units: bright and dark terrains. These two main units appear also in the Huygens landing site images acquired by the DISR instrument (**Figure 1 and Tomasko *et al.*, 2005**).

One of the most striking morphological structures in the 2.03 µm image is a circular feature located at the northern edge of the western large bright feature (centered at 159°E longitude and -7° latitude). This small circular feature presents four pixels at its centre, bright in all images. The bright centre is surrounded by a ring of darker material. This structure also appears in most of the band ratios. The dark ring has a diameter of almost 150 km and may correspond to an impact crater with a central peak (see **Figure 8**).

[Figure 8]

Other features that were not detectable in the non corrected ratios (**Figure 6**) show up in the ones which have been subtracted for the atmospheric aerosol contribution (**Figure 7**).

In particular, a spatially well-delimited dark area (unit 1 in **Figure 7**) appears at the center of the 1.59/1.27 µm image just northeast of the large bright structure. This feature may correspond to a region dominated by an absorber at 1.59 µm, not present or less abundant elsewhere in the scene. This anomalous dark area, completely disconnected from albedo variations on single band images, is also observed with the same spatial extension in all corrected ratios involving the 1.59 and 2.03 µm bands, but not the 1.08 and 1.27 µm. It could indicate a spatially delimited presence of a constituent that absorbs at 1.59 µm and at 2.03 µm, and much less at 1.08 and 1.27 µm bands. The dark ringed interior of the small "crater" seems to be constituted with the same material and may witness the presence of deposits of material different in nature than that of the surroundings. It should be noted that the same dark diffuse areas located at the frontier between bright and dark regions can be observed in the T5 VIMS ratio images. Similar dark regions appearing in band ratios seem to be also related to fuzzy areas observed at the frontiers between albedo units seen in images of the ISS/Cassini camera (**Figure 6 in Porco *et al.*, 2005**). The peculiar dark feature seen on VIMS ratio images seems to extend into the area where the Huygens probe landed, just north of the east ending of the large central bright patch (see **Figure 2**). Here the DISR/Huygens spectral measurements show evidence for reddening of the high I/F area with respect to low I/F area (**Tomasko *et al.*, 2005**). The landing site presents an 827/751 nm ratio lower than the same ratio in the bright region just south of it, similar to what we see with VIMS in the 1.59/1.08 μm ratio.

A relatively bright feature (unit 2 in **Figure 7**) appears at the northeast corner of the 1.59/1.08 μm and 2.79/1.08 μm ratio images and at the northeast and southeast corner of the 1.59/1.27 μm and 2.03/1.27 μm ratio images. It is related to the darkest units we can find in the single band images. That suggests another local exposure of a material with a different spectral behaviour, more reflective at 1.59 and 2.03 μm than at 1.27 μm.

Two small strikes, elongated westward, can also be seen on the 2.03 μm image at the centre of the north-east dark unit (unit 3 in **Figure 7**). These structures are distinguishable in all the ratio images, except those involving wavelengths where the flux is very low (2.79 and 5 μm – refer to the last row of the **Figure 7**), and appear bright at all wavelengths.

Apart from unit 1, 2 and 3, the other regions are globally homogeneous and do not exhibit significant spectral differences in all the ratio images involving the 1.08, 1.27, 1.59 and 2.03 μm bands.

**[Figure 9]**

**Figure 9** shows the same image ratios as **Figure 7**, but for the TB observations (thus 48 days after TA encounter). It is particularly interesting to see that the same heterogeneities appear in these images, which suggest that the observed variations are linked to the surface and that no major change in morphology has occurred during this time span.

Finally, the 5/2.79 µm ratio presents a particularly bright area south of the large western bright structure (unit 4 in **Figure 7**: it is 1.5 to 2 times brighter than the surrounding terrains). This area does not appear distinctively on other ratio images and may indicate the presence of materials in a slightly different physical state only distinguishable at longer wavelengths: this region may be warmer than the surroundings or composed of thinner grain size ices witnessing fresh deposits (for example, water frost is more reflective around 5 µm than coarse grain water ice). Unfortunately, this area was not imaged during the TB flyby, which prevents the study of its evolution with time.

*4.3. First implications for surface composition.*

Band ratios can give us clues for the spectral behaviour of some particular units within the Huygens site images. They can be compared to some candidate molecules thought to be present at the surface of Titan and to the spectrum of Titan's surface locally recorded by DISR (**Tomasko *et al.*, 2005**).

Since the spectral region around 2.6-2.8 µm corresponds to the beginning of a strong water ice absorption band, the 2.79/2.69 µm ratio should be very sensitive to the presence of pure water ice expo-sure (**Figure 10**). If pure water frost (*i.e.* with grains finer than few tens of microns) is present at the Huygens landing site, this should make the ratio of the 2.79 to the 2.69 µm image less than one, which is never the case. That suggests that pure water ice is not dominantly exposed around the Huygens landing site. VIMS high-resolution observations during TA led to same conclusion (**Sotin *et al.*, 2005**). If water ice is present, the slight variations within 2.79/2.69 µm ratio image indicate that bright and dark regions do not differ markedly in their water ice content. It is important to note that VIMS is only sensitive to the first few tens of microns of the surface, and that a coating of few millimeters is enough to mask the spectral signature of the underlying materials. Since water is likely to be a major constituent of Titan's surface, it is probably largely mixed with, or covered with darker photochemical material as suggested by **Griffith *et al.* (2003)** and in this case partly inaccessible to VIMS. Alternatively, a grain size effect could explain the lack of variations in the 2.79/2.69 µm ratio if water ice of coarse grains is present. Indeed, water ice with coarse grains (more than 1 mm) still displays broad 1.5 and 2 µm bands, but the 2.6-2.8 µm drop-off is much less pronounced than in the case of fine grained water ice.

[Figure 10]

Enrichment in $H_2O$ ice, either in a hydrate state or mixed with a material that globally darkens its spectrum (as tholins) is a good candidate to explain the spectral behaviour of the dark unit 1. This molecule has absorption bands at 1.6 and 2 µm, particularly strong and broad in the case of large grained $H_2O$ ice (**Figure 10**), and comparatively higher reflectance in the spectral region between 1 and 1.3 µm. Other components likely to be present at Titan's surface, as icy $CO_2$ and $CH_4$, do not fulfil simultaneously all the conditions to match the observed spectral behaviour of this "dark spot". Pure $CO_2$ ice exhibits a strong absorption band at 2 µm but does not absorb much in the 1.6 µm region (**Figure**

**10**), which is not compatible with the ratios with the VIMS 1.59 µm band. On the other hand, $CH_4$ ice is much too reflective in all the atmospheric windows to be responsible for the observed dark patch.

At this time, no candidate was identified to explain the anomalous behaviour of unit 2. The spectral behaviour of this region may be compatible with local depletion in $H_2O$ ice or coating by a component that globally raises the spectral reflectance at 1.59 and 2.03 µm with respect to the one at 1.27 µm.

## 5. Conclusions

VIMS has provided the first context images of the Huygens descent and entry site, with a resolution ranging from 16 to 14.4 km/pixel. We have discovered a circular structure almost 150 km in diameter which could correspond to an impact crater with a central peak. The main variations in VIMS images are dominated by bright and dark terrains with no major spectral differences between the two units. An empirical correction for the atmospheric scattering effects, validated by a first order theoretical modelling, still reveals subtle spectral variations in ratio images. In particular, it shows an anomalous diffuse dark area in the contact between bright and dark regions, with an increased absorption at 1.59 and 2.03 µm. This feature could be consistent with a local enrichment in water ice content. Altogether we have demonstrated that the joint use of ratios between images corrected from haze scattering involving bands at 1.27, 1.59 and 2.03 µm can provide a good indicator of the relative content of water ice.

The ratio images of the Huygens landing site used in this study allowed us to identify a number of surface units with specific spectral signatures, even if it is still too soon to identify without ambiguities the molecules that may be related to these heterogeneities. However, it is possible to draw some hypotheses about the nature of these "spectral" units and to link the detected heterogeneities with the geological history of this site.

**[Figure 11]**

**Figure 11** summarizes our first interpretations for the "spectro-"morphological properties we proposed for the Huygens landing area. We decomposed the site into 7 units: three main units, associated with albedo variations (bright, dark and diffuse dark areas, respectively light yellow, brown and light brown on the map), and four, associated with heterogeneities observed in the ratio images, totally uncorrelated with albedo (units 1, 2, 3 and 4 identified in the **section 3.2**, respectively dark blue, cyan, yellow and greenish blue on the map). It is tempting to associate the bright area to elevated terrains. Although we have not carried out a detailed study of the correlation between albedo and altitude, the few differences observed between spectra of the dark and bright areas may reveal a difference in the depth of the $CH_4$ column density and therefore the elevation of the terrains. Future data by the radar and finer methane absorption modelling will be useful to calibrate the albedo variations against topography. The dark structure that has relative absorption signatures at 1.59 and 2.03 μm (unit in blue in **Figure 11**) extends to the lower plains where small round pebbles were imaged by the Huygens probe. These blocks which could have been washed out and carried by ancient flows of liquid methane from higher lands are mainly composed of water ice (**Tomasko et al., 2005**). They could account for the slight areal enrichment in water ice that may be detected from orbit in VIMS spectra.

A detailed atmospheric model is required in order to accurately correct VIMS images from the complete atmospheric contribution. Spectra acquired by the Huygens DISR instrument directly on the surface will be very useful in providing a ground truth to test this atmospheric correction process. VIMS images acquired during the 4 year tour of the CASSINI spacecraft will provide opportunities to generalize the findings of Huygens in a specific area. The radar T8 observation of the Huygens landing site acquired in October 2005 can also give complementary information. Similarly, it would be extremely valuable to have other VIMS images of this site, but with a better spatial resolution. A 1-2 km resolution, which is needed for a geomorphological analysis, is achievable only if VIMS observes at closest approach.

**Acknowledgements.**

We thank J-P Combe for his help with data reprojection, G. Tobie for very useful discussions, and B. Schmitt for the spectra of ices he provided. The work by the Cassini team and the dedication of the VIMS team for the design of the observations is much appreciated. This work is supported by CNES and CNRS.

**References.**

Brown, R.H., Baines, K.H., Bellucci, G., Bibring, J.-P., Buratti, B.J., Capaccioni, F., Cerroni, P., Clark, R.N., Coradini, A., Cruikshank, D.P., Drossart, P., Formisano, V., Jaumann, R., Langevin, Y., Matson, D.L., McCord, T.B., Mennella, V., Nelson, R.M., Nicholson, P.D., Sicardy, B., Sotin, C., Amici, S., Chamberlain, M.A., Filacchione, G., Hansen, G., Hibbitts, K., Showalter, M., 2003. Observations with the Visual and Infrared Mapping Spectrometer (VIMS) during Cassini's flyby of Jupiter. *Icarus* **164**, 461-470.

Cabane, M., Chassefiere, E., Israel, G., 1992. Formation and growth of photochemical aerosols in Titan's atmosphere. *Icarus* **96**, 176-189.

Combes, M., Cara, C., Drossart, P., Encrenaz, T., Lellouch, E., Rosenqvist, J., Bibring, J.-P., Erard, S., Gondet, B., Langevin, Y., Soufflot, A., Moroz, V. I., Grygoriev, A. V., Ksanfomality, Lu. V., Nikolsky, Y. V., Sanko, N. F., Titov, D. V., Forni, O., Masson, P., Sotin, C., 1991. Martian atmosphere studies from the ISM experiment. *P&SS* **39**, 189-197.

Coustenis, A., Lellouch, E., Maillard, J.P., McKay, C.P., 1995. Titan surface: composition and variability from its near-infrared albedo. *Icarus* **118**, 87–104.

Coustenis, A., Lellouch, E., Combes, M., Wittemberg, R., McKay, C.P., Maillard, J.P., 1997. Titan's atmosphere and surface from infrared spectroscopy and imaging. In: Cosmovici, C.B., Bowyer, S., Werthimer, D. (Eds.), Astronomical and Biochemical Origins and the Search for Life in the Universe. In: *Proceedings of IAU Colloquium*, Vol. **161**, 277–284.

Coustenis, A., Gendron, E., Lai, O., Véran, J.P., Woillez, J., Combes, M., Vapillon, L., Fusco, T., Mugnier, L., Rannou, P., 2001. Images of Titan at 1.3 and 1.6 µm with adaptative optics at the CFHT. *Icarus* **154**, 501–515.

Coustenis, A., Negrão, A., Salama, A., Schulz, B., Lellouch, E., Rannou, P., Drossart, P., Encrenaz, T., Schmitt, B., Boudon, V., Nikitin, A., 2005. Titan's 3-micron spectral region from ISO high-resolution spectroscopy. *Icarus,* in press .

Griffith, C.A., Owen, T., Wagener, R., 1991. Titan's surface and troposphere, investigated with ground-based, near-infrared observations. *Icarus* **93**, 362-378.

Griffith C.A., Owen, T., Geballe, T.R., Rayner, J., Rannou, P., 2003. Evidence for the exposure of water ice on Titan's surface. *Science* **300**, 628–630.

Khare, B.N., Sagan, C., Arakawa, E.T., Suits, F., Callcott, T.A., Williams, M.W., 1984. Optical constants of organic tholins produced in a simulated titanian atmosphere: from soft X-ray to microwave frequencies. *Icarus* **60**, 127-137.

Ishimaru, A., 1997. Scattering and absorption of a wave by a single particle. In: Dudley, D.G. (Ed.), Wave propagation and scattering in random media. Oxford University Press, 9-40.

Langevin, Y., Poulet, F., Bibring, J.P., Schmitt, B., Douté, S., Gondet, B., 2005. Summer evolution of the north polar cap of Mars as observed by OMEGA/Mars Express. *Science* **307**, 1581-1584.

Lara, L.M., Lellouch, E., López-Moreno, J.J., Rodrigo, R., 1996. Vertical distribution of Titan's atmospheric neutral constituents. *J. Geophys. Res.* **101** (E10), 23261–23284.

Lara, L.M., Lellouch, E., López-Moreno, J.J., Rodrigo, R., 1998. Erratum: Vertical distribution of Titan's atmospheric neutral constituents. *J. Geophys. Res.* **103** (E11), 25775–25776.

Le Mouélic, S., Langevin, Y., Erard, S., 1999. A new reduction approach for the Clementine near infrared data set. Application to Aristillus, Aristarchus and Kepler. *J. Geophys. Res.*, **104**, 3833-3844.

McKay, C.P., Pollack, J.B., Courtin, R., 1989. The thermal structure of Titan's atmosphere. *Icarus* **80**, 23-53.

Muhleman, D.O., Grossman, A.W., Butler, B.J., Slade, M.A., 1990. Radar reflectivity of Titan. *Science* **248**, 975–980.

Porco, C.C., Baker, E., Barbara, J., Beurle, K., Brahic, A., Burns, J.A., Charnoz, S., Cooper, N., Dawson, D.D., Del Genio, A.D., Denk, T., Dones, L., Dyudina, U., Evans, M.W.; Fussner, S., Giese, B., Grazier, K., Helfenstein, P., Ingersoll, A.P., Jacobson, R.A., Johnson, T.V., McEwen, A., Murray, C.D.; Neukum, G., Owen, W.M., Perry, J., Roatsch, T., Spitale, J., Squyres, S., Thomas, P., Tiscareno, M., Turtle, E.P., Vasavada, A.R., Veverka, J., Wagner, R., West, R., 2005. Imaging of Titan from the Cassini spacecraft. *Nature* **434**, Issue 7030, 159-168.

Quirico, E., Schmitt, B., 1997. Near-infrared spectroscopy of simple hydrocarbons and carbon oxides diluted in solid $N_2$ and as pure ices: Implications for Triton and Pluto. *Icarus* **127**, 354-378.

Rages, K., Pollack, J.B., 1983. Vertical distribution of scattering haze in Titan's upper atmosphere. *Icarus* **55**, 50-62.

Rodriguez, S., Paillou, P., Dobrijevic, M., Ruffié, G., Coll, P., Bernard, J. M., Encrenaz, P, 2003. Impact of aerosols present in Titan's atmosphere on the CASSINI radar experiment. *Icarus,* **164**, 213-227.

Sobolev, V.V., 1975. Light scattering in planetary atmospheres. Pergamon Press, Oxford.

Sotin, C., Jaumann, R., Buratti, B. J., Brown, R. H., Clark, R. N., Soderblom, L. A., Baines, K. H., Bellucci, G., Bibring, J.-P., Capaccioni, F., Cerroni, P., Combes, M., Coradini, A., Cruikshank, D. P., Drossart, P., Formisano, V., Langevin, Y., Matson, D. L., McCord, T. B., Nelson, R. M., Nicholson, P. D., Sicardy, B., Le Mouélic, S., Rodriguez, S., Stephan, K., Scholz, C. K., 2005. Release of volatiles from a possible cryovolcano from near-infrared imaging of Titan. *Nature* **435**, Issue 7043, 786-789.

Smith, P.H., Lemmon, M.T., Lorenz, R.D., Sromovsky, L.A., Caldwell, J.J., Allison, M.D., 1996. Titan's surface, revealed by HST imaging. *Icarus* **119**, 336-349.

Tomasko, M. G., Buchhauser, D., Bushroe, M., Dafoe, L. E., Doose, L. R., Eibl, A., Fellows, C., Farlane, E. M., Prout, G. M., Pringle, M. J., Rizk, B., See, C., Smith, P. H., Tsetsenekos, K., 2002. The Descent Imager/Spectral Radiometer (DISR) experiment on the Huygens entry probe of Titan. *Space Sci. Rev.* **104**, 469–551.

Tomasko, M.G., Archinal, B., Becker, T., Bézard, B., Bushroe, M., Combes, M., Cook, D., Coustenis, A., de Bergh, C., Dafoe, L. E., Doose, L., Douté, S., Eibl, A., Engel, S., Gliem, F., Grieger, B., Holso, K., Howington-Kraus, E., Karkoschka, E., Keller, H. U., Kirk, R., Kramm, R., Küppers, M., Lanagan, P., Lellouch, E., Lemmon, M., Lunine, J., McFarlane, E., Moores, J., Prout, G. M., Rizk, B., Rosiek, M., Rueffer, P., Schröder, S. E., Schmitt, B., See, C., Smith, P., Soderblom, L., Thomas, N., West, R., 2005. The Descent Imager/Spectral Radiometer (DISR) Experiment on the Huygens entry probe of Titan. *Nature,* **438**, Issue 7069, 765-778.

Toublanc, D., Parisot, J.P., Brillet, J., Gautier, D., Raulin, F., McKay, C.P, 1995. Photochemical modeling of Titan's atmosphere. *Icarus* **113**, 2–26.

Yung, Y.L., Allen, M., Pinto, J.P., 1984. Photochemistry of the atmosphere of Titan: comparison between model and observations. *Astrophys. J.* **55**, 465–506.

**Captions:**

**Table 1:** *Comparison between empirical and theoretical correction of the atmospheric scattering contribution.*

**Figure 1: (a):** *This composite image was produced by the DISR instrument onboard the ESA's Huygens probe, on 14$^{th}$ January 2005, during the descent to Titan. It shows a full 360-degree view around Huygens (medium-altitude (17 to 8 km) panoramic mosaic projected from 8 km). The mosaic shows a boundary between light and dark areas. As the probe descended, it drifted over a bright plateau (left side of the image) and was heading towards its landing site in a dark area (symbolized by the grey circle at the center of the image). Narrow dark lineaments, interpreted as channels which might still contain liquid material, cut brighter terrain* (**Tomasko et al., 2005**). **(b):** *A view of Titan from the VIMS instrument on the Cassini orbiter. The Huygens probe landed in the small white circle on the boundary of the bright and dark regions. The size of the circle shows the field of view of the Huygens DISR imager from an altitude of 20 kilometres* (© ESA/NASA/JPL/University of Arizona).

**Figure 2:** *(a)***:** *VIMS global coverage of Titan's surface obtained after the first height Cassini flybys. The red rectangle represents the location of TA and TB observations of the Huygens landing site.* **(b)** *and* **(c)** *are respectively the TA raw images for the long and short time exposures.* **(d)***: TB raw 10 cubes mosaic. The mean spatial resolution of the images (b), (c) and (d) is around 15 km/pixel.* **(e)** *and* **(f)***: same as (b) and (c) but reprojected on a Titan cylindrical map.* **(g)***: TB reprojected mosaic. At these resolutions, no major variation is observed between the TA and TB flybys. The red circle on each reprojected image indicates the estimated Huygens landing site (Latitude: -10,3°, Longitude: 167,6°E -* **Tomasko et al., 2005**).

**Figure 3:** *I/F contours of the TB 2.03μm mosaic (**Figure 2g** and reproduced here at the lower left) overlaying the TA 2.03 μm image (**Figures 2f** and reproduced here at the upper left). Contours of the TB mosaic were chosen in order to delimit the main I/F units that appear in common on the 2.03 μm TA and TB images. Same dynamic range was fixed for TA and TB images to allow a rigorous comparison. The contours of the TB image match the I/F variations observed at TA, indicating that at this spatial resolution (~15 km/pixel) no major change occurred during the 48 days separating the two encounters.*

**Figure 4:** *(a): The two most different I/F spectra extracted from the long time exposure VIMS cube of Huygens landing site, acquired during the TA flyby. (b): zoom in the 0.9-2.2 μm range. These spectra (corresponding to the black and grey rectangles on the **right** image) are representative of bright (dotted line) and dark terrains (black plain line). The seven atmospheric windows are labelled on Figure 4a with their respective spectral position. (c): ratio between the two spectra.*

**Figure 5:** *Additive contribution of the scattering by aerosols to the VIMS I/F observations of Titan. The plain and dash-point black lines correspond to the results of the empirical method described in the text applied to TA and TB VIMS images, respectively. The associated theoretical curves (plain and dash-point red lines) were obtained by modelling the reflection of solar light on a unique aerosol layer with the following characteristics: a vertical extension $H = 200\,km$, composed of particles with a radius $r = 0.18\,\mu m$, homogeneously distributed with a number density $\rho = 100\,cm^{-3}$. The empirical and theoretical approaches provide consistent results.*

**Figure 6:** *Long time exposure georeferenced images at 1.08 and 2.03 μm of the Huygens landing site acquired by VIMS during TA flyby (north is up). Ratio images using the bands centered on atmos-*

*pheric windows are presented in order to emphasize possible regional variations in composition. Band ratios using short wavelengths are correlated to the albedo due to a strong global additive contribution coming from the light scattered in Titan's atmosphere. The color bars indicate the values of the ratio within each ratio image.*

**Figure 7:** *Same figure as **Figure 6** but with ratio images empirically corrected from atmospheric haze scattering. As expected, the correlation with albedo has been significantly reduced. Local heterogeneities disconnected from albedo variations appear in the ratio images (see in particular the dark diffuse area at the center of all ratios involving the 1.59 and 2.03 µm images). As in **Figure 6**, the color bars indicate the values of the ratio within each ratio image. The four main units which are discussed in the text are displayed on the relevant ratio images.*

**Figure 8:** *Possible detection of an impact crater (~150 km in diameter) in the vicinity of the Huygens landing site (longitude: 159°E, latitude: -7°).* **Right:** *Qualitative estimation of the "crater" topography based on the assumption that I/F level in the image (corrected from additive aerosol contribution) is only correlated to the local terrain slope. The viewpoint was selected eastward, along the cross-sun direction during the observation. The same method was used to derive local topography on VIMS high-resolution images (**Sotin et al., 2005**). The retrieved topography shown in the right panel was vertically exaggerated in order to enhance the crater-like shape of the selected area. The estimated depth of the hypothesized crater basin is only a few hundreds of meters.*

**Figure 9:** *Same as Figure 7, but for the TB flyby (13 December 2004). The same heterogeneities as the ones observed at TA are detected, which suggest that the observed variations are linked to the surface and not to transient atmospheric phenomena.*

**Figure 10:** *Simulated reflectance spectra of candidate molecules for Titan surface composition discussed in the text: black plain line for coarse grain water ice and grey dotted line for fine grain water ice* (**Langevin *et al.*, 2005**)*, and grey plain line for $CO_2$ ice [B. Schmitt, private communication].*

**Figure 11:** *Interpretation of the landing site observations. On the basis of albedo and spectral variations in ratio images, a geological interpretation of the Huygens landing site has been obtained. The geological map is drawn using criteria defined in the text.*

**Table 1**

| Atmospheric window (μm) | Empirical method | | | | Modelling | | |
| --- | --- | --- | --- | --- | --- | --- | --- |
| | TA flyby | | TB flyby | | | | |
| | Additive contribution to the band I/F - $A_\lambda$ | % of mean I/F over the TA image | Additive contribution to the band I/F - $A_\lambda$ | % of mean I/F over the TB image | Additive contribution to the band I/F | % of mean I/F over the TA image | Aerosol layer transmittance ($e^{-\tau}$) |
| 1.08 | 0.138 | 80 % | 0.134 | 81 % | 0.131 | 75 % | 0.13 |
| 1.27 | 0.098 | 72 % | 0.088 | 68 % | 0.099 | 73 % | 0.35 |
| 1.59 | 0.057 | 61 % | 0.05 | 58 % | 0.055 | 59 % | 0.65 |
| 2.03 | 0.028 | 40 % | 0.02 | 31 % | 0.023 | 39 % | 0.84 |
| 2.75 { 2.69 | 0.012 | **66 %**[a] | 0.011 | **64 %**[a] | 0.007 | 39 % | 0.94 |
| 2.79 | 0.012 | **57 %**[a] | 0.011 | **56 %**[a] | 0.0055 | 26 % | 0.87 |
| 5 | 0 | -- | 0 | -- | 0.0004 | 2 % | 0.98 |

[a] *anomalous high values for the additive scattering component inferred from the empirical method described in the text. A possible explanation is that the 2.75 μm region has very low I/F levels compared with the others spectral windows. Our empirical method overestimates in an arbitrary way the additive constant needed to minimize image ratios in this spectral region. The corresponding $A_\lambda$ values for the 2.69 and 2.79 μm bands must therefore be considered with care.*

**Figure 1**

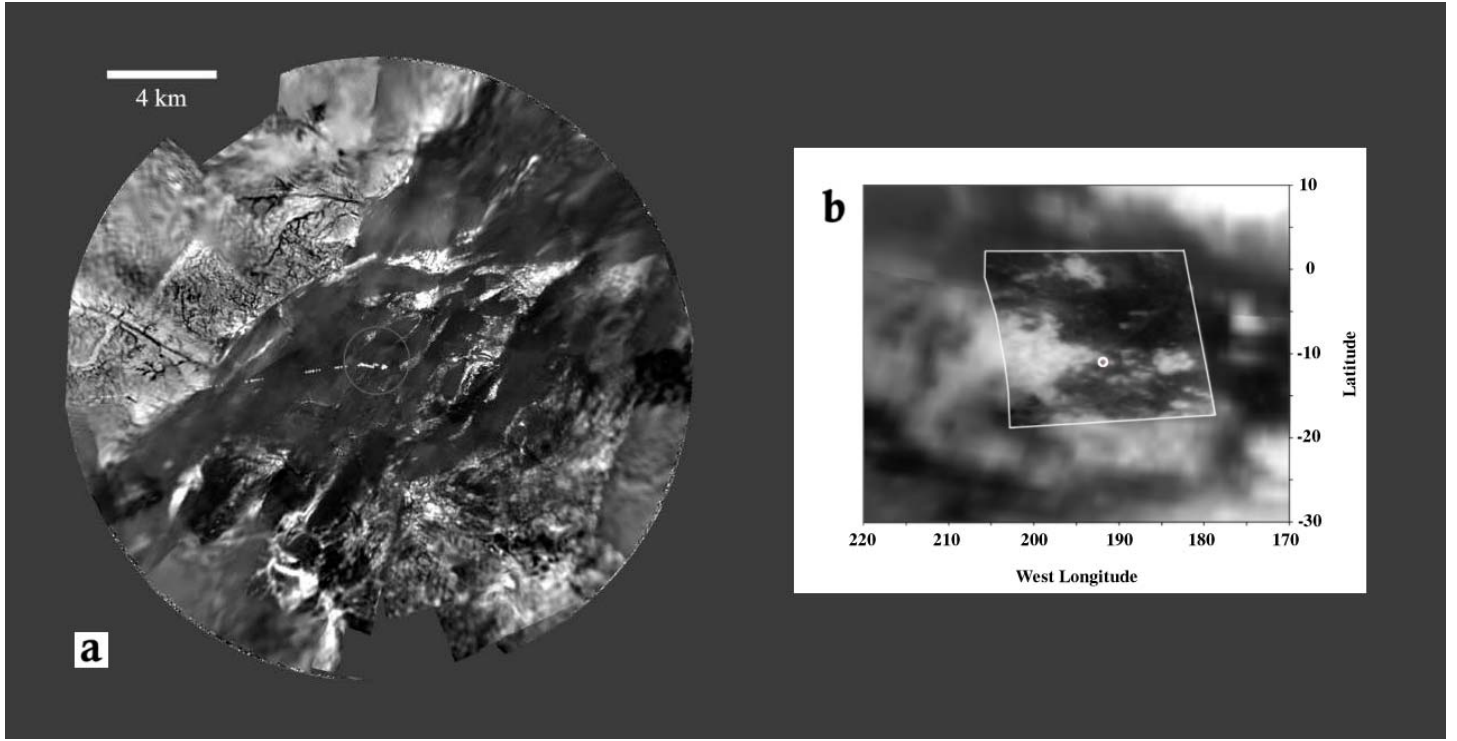

**Figure 1**

**Figure 2**

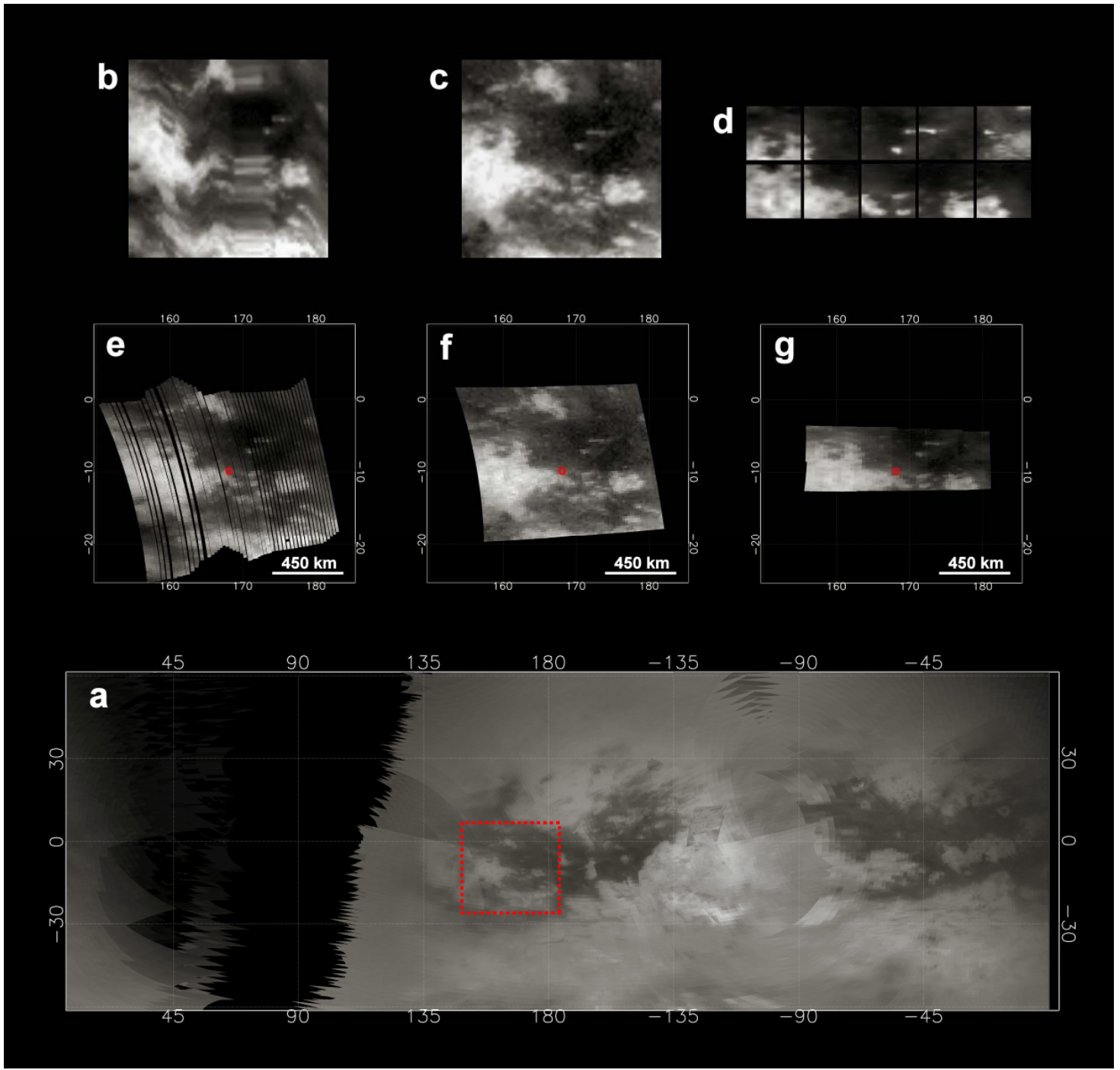

**Figure 3**

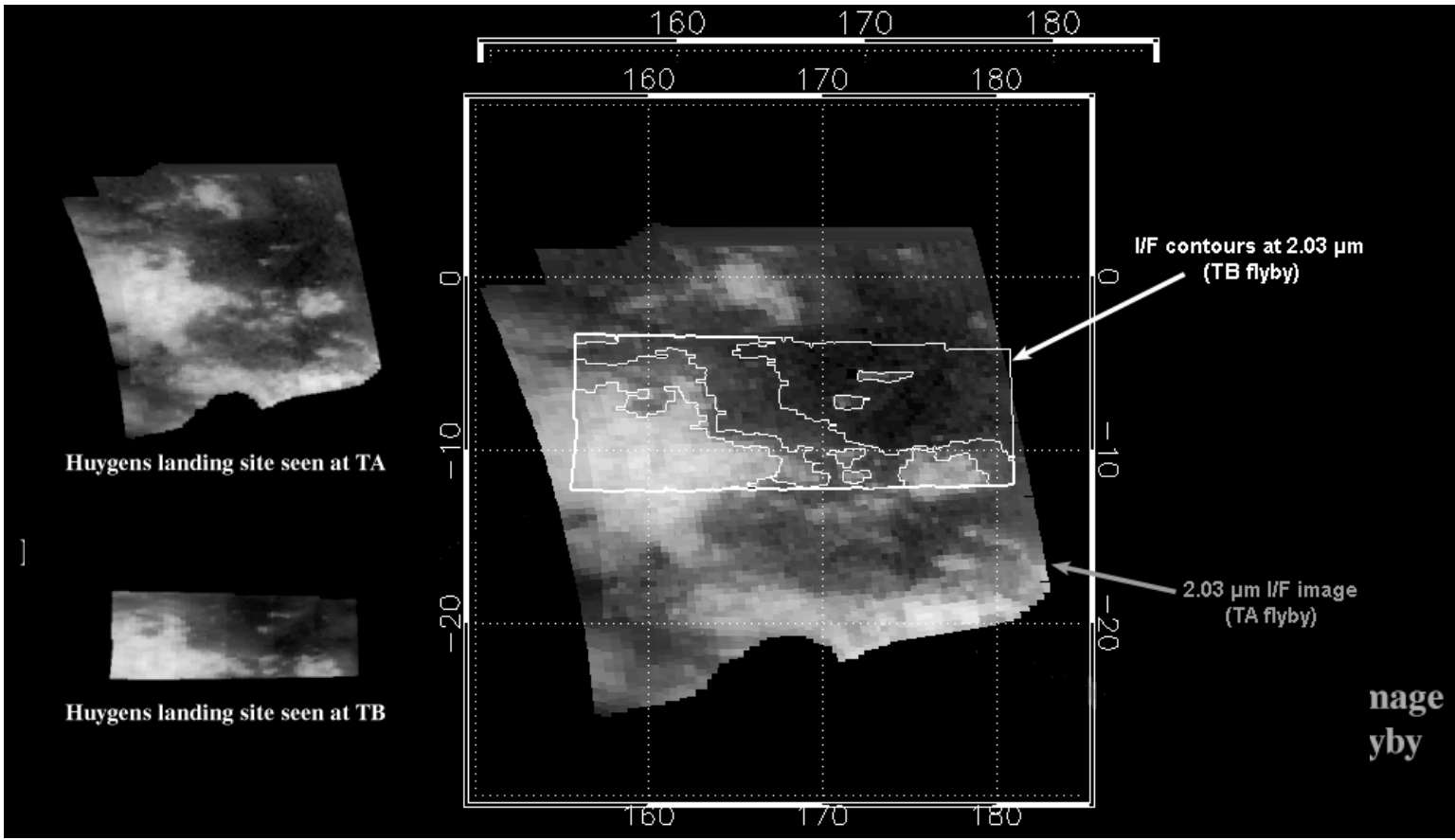

**Figures 4ab and 4c**

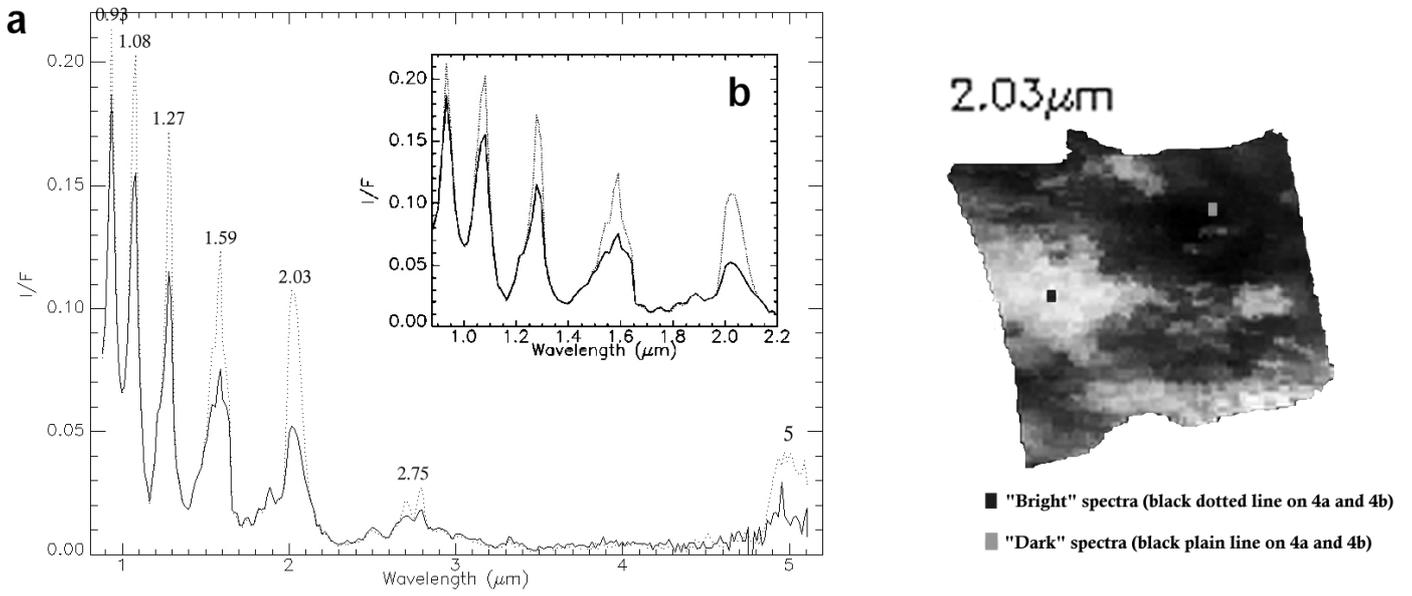

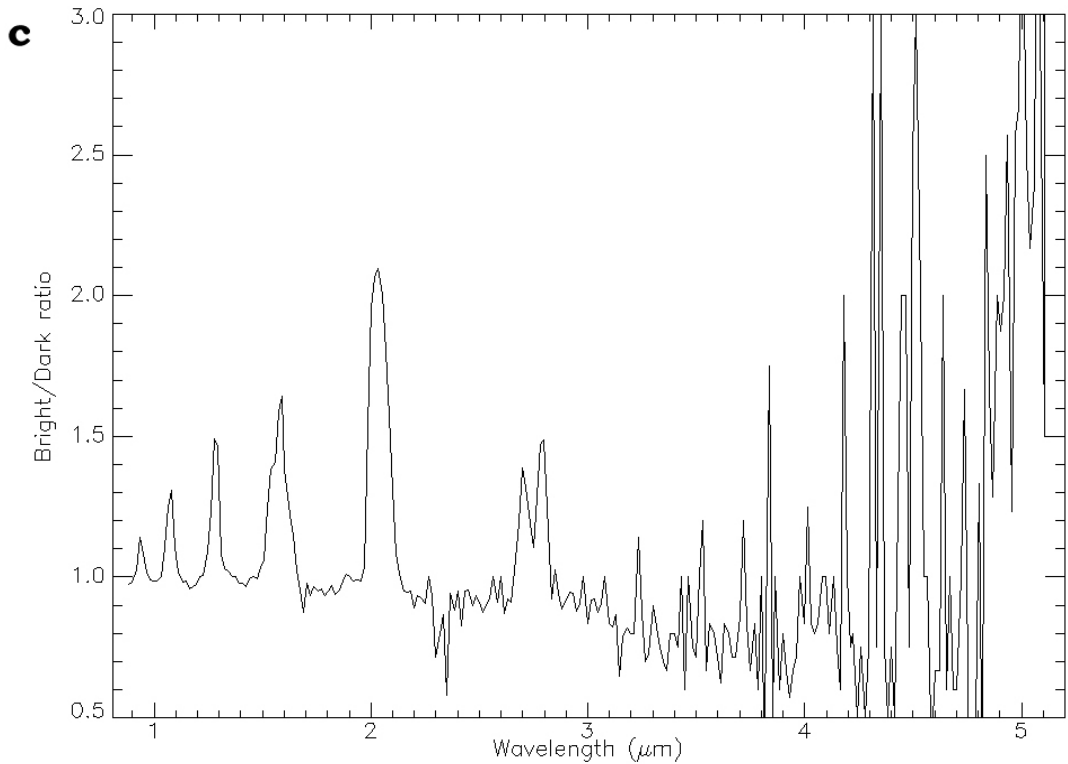

**Figure 5**

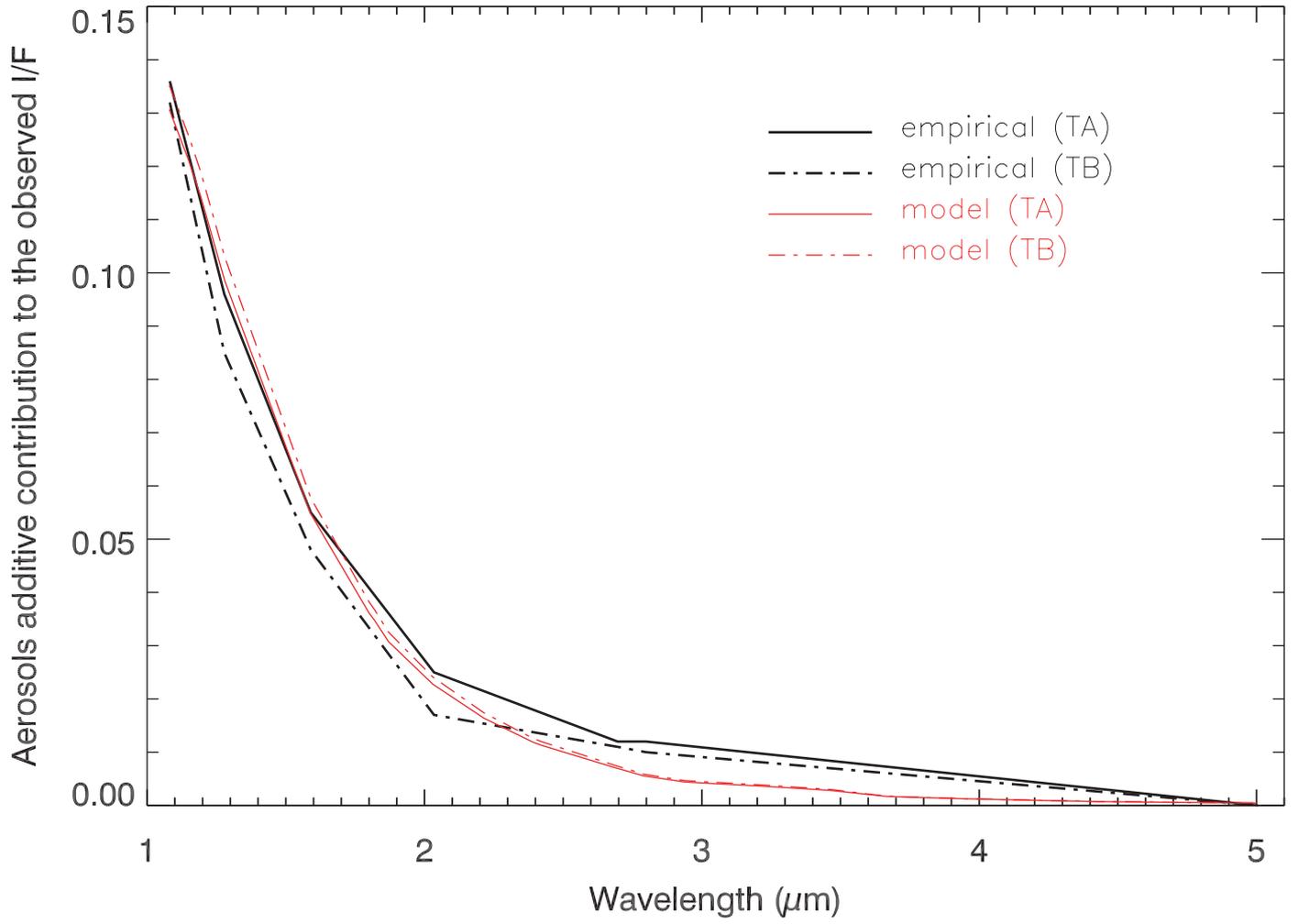

**Figure 6**

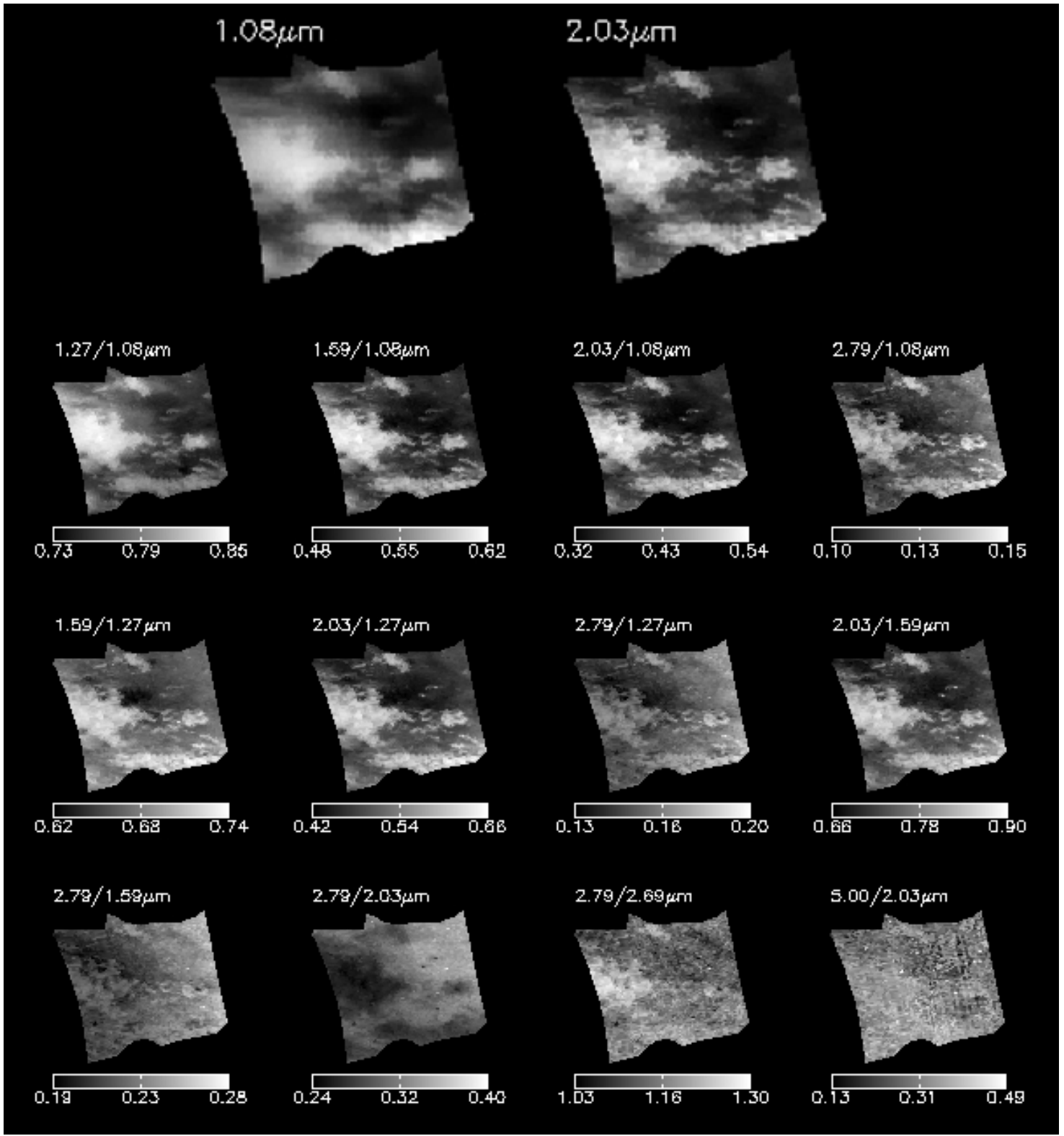

**Figure 7**

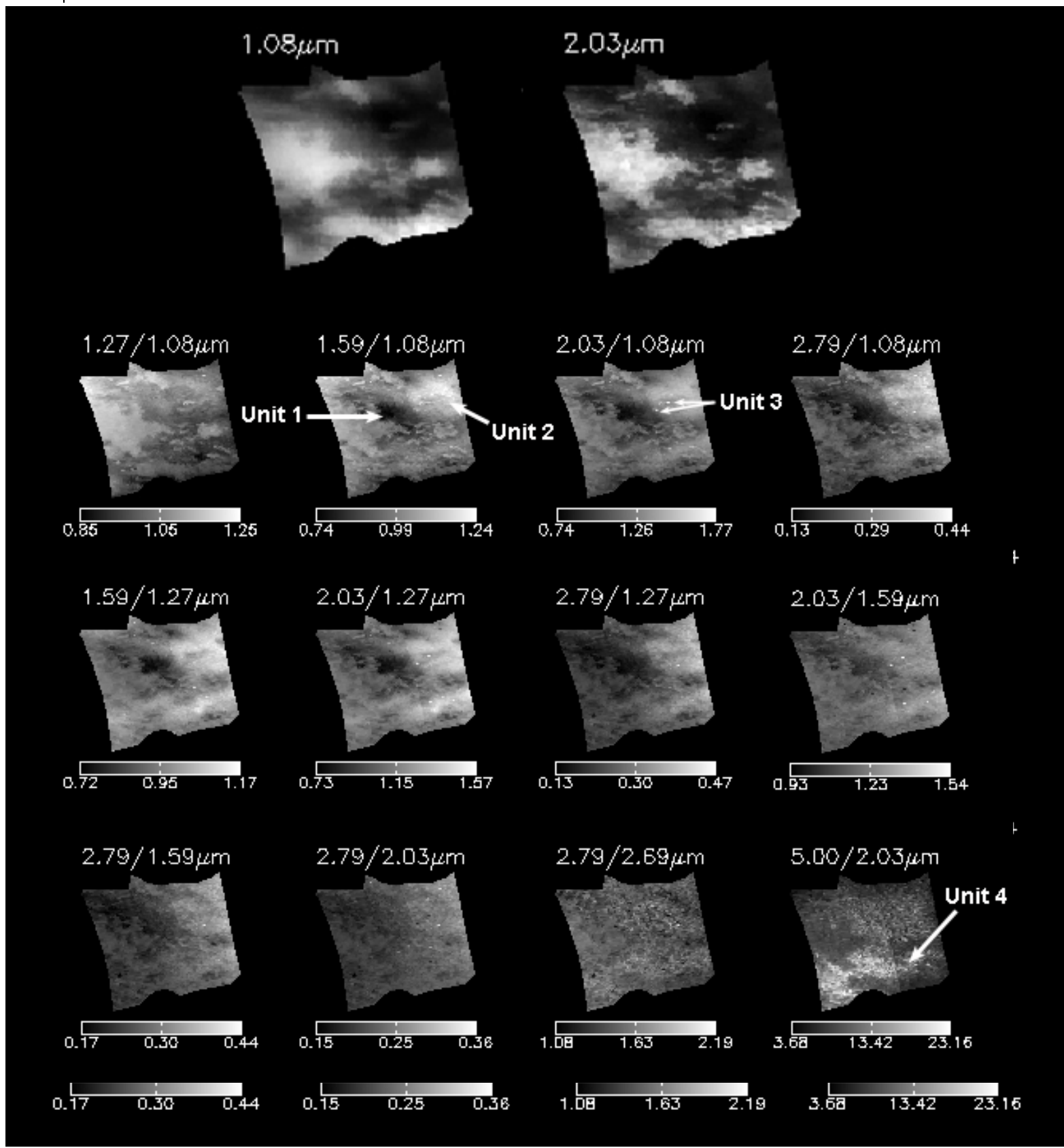

**Figure 8**

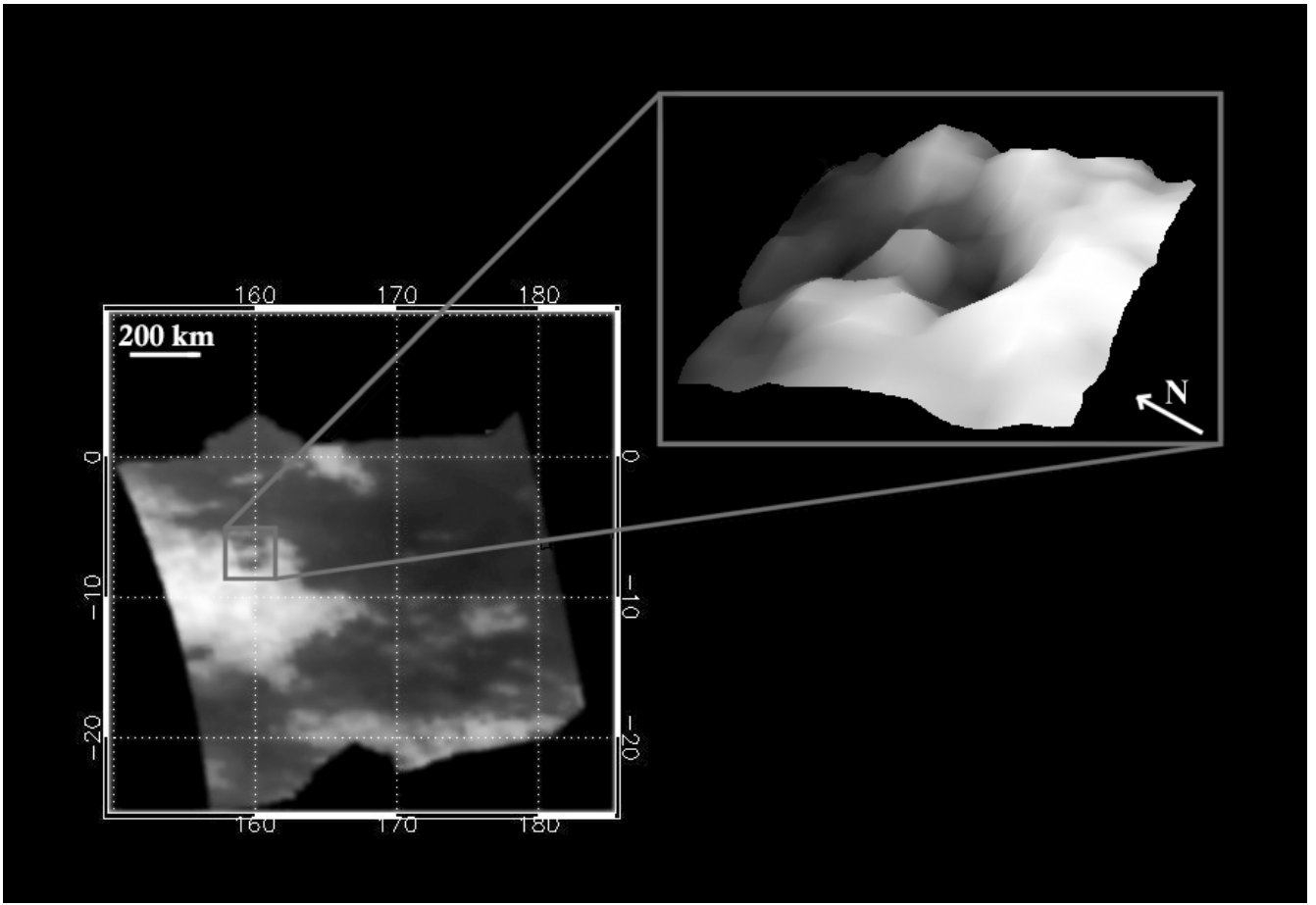

**Figure 9**

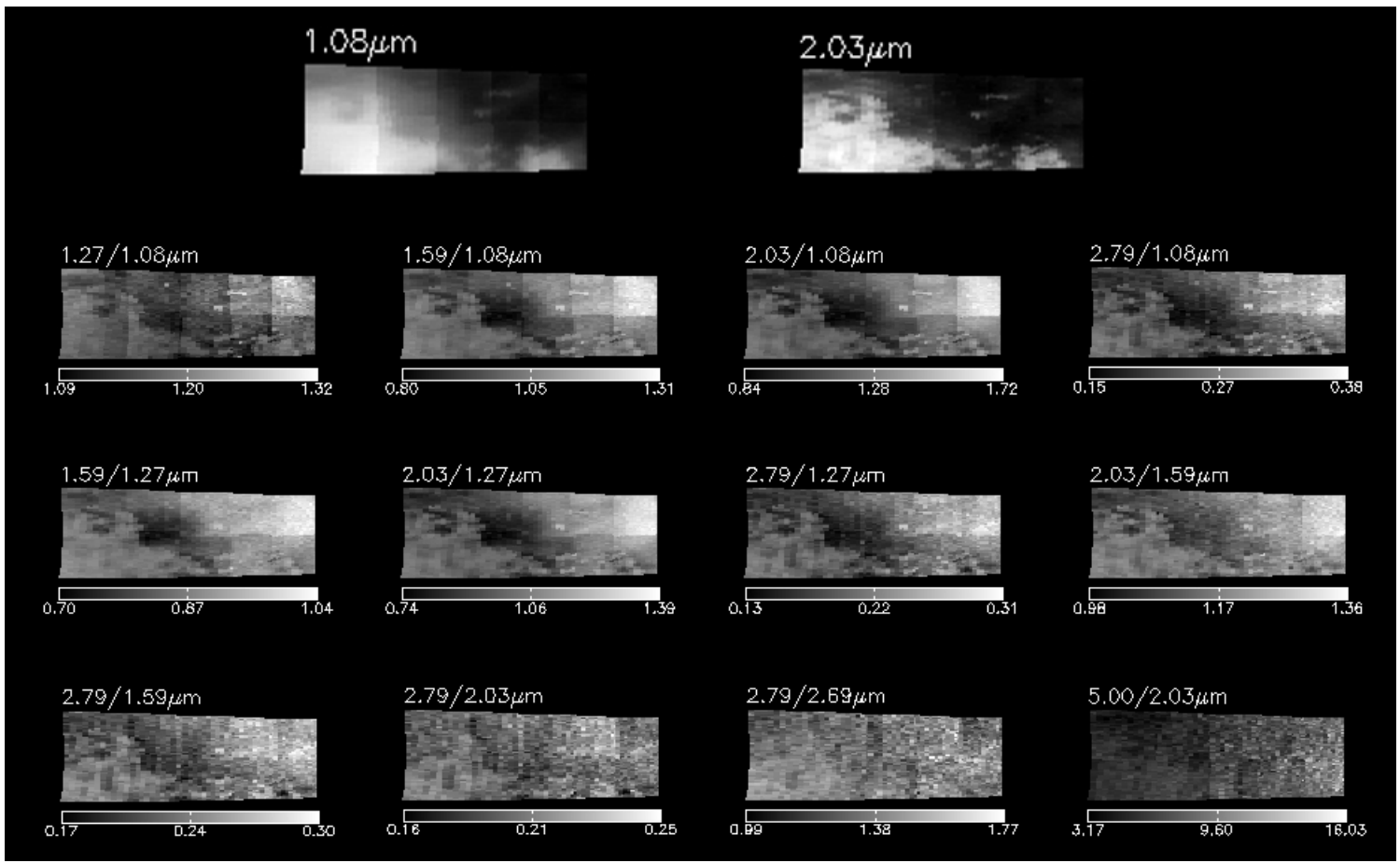

**Figure 10**

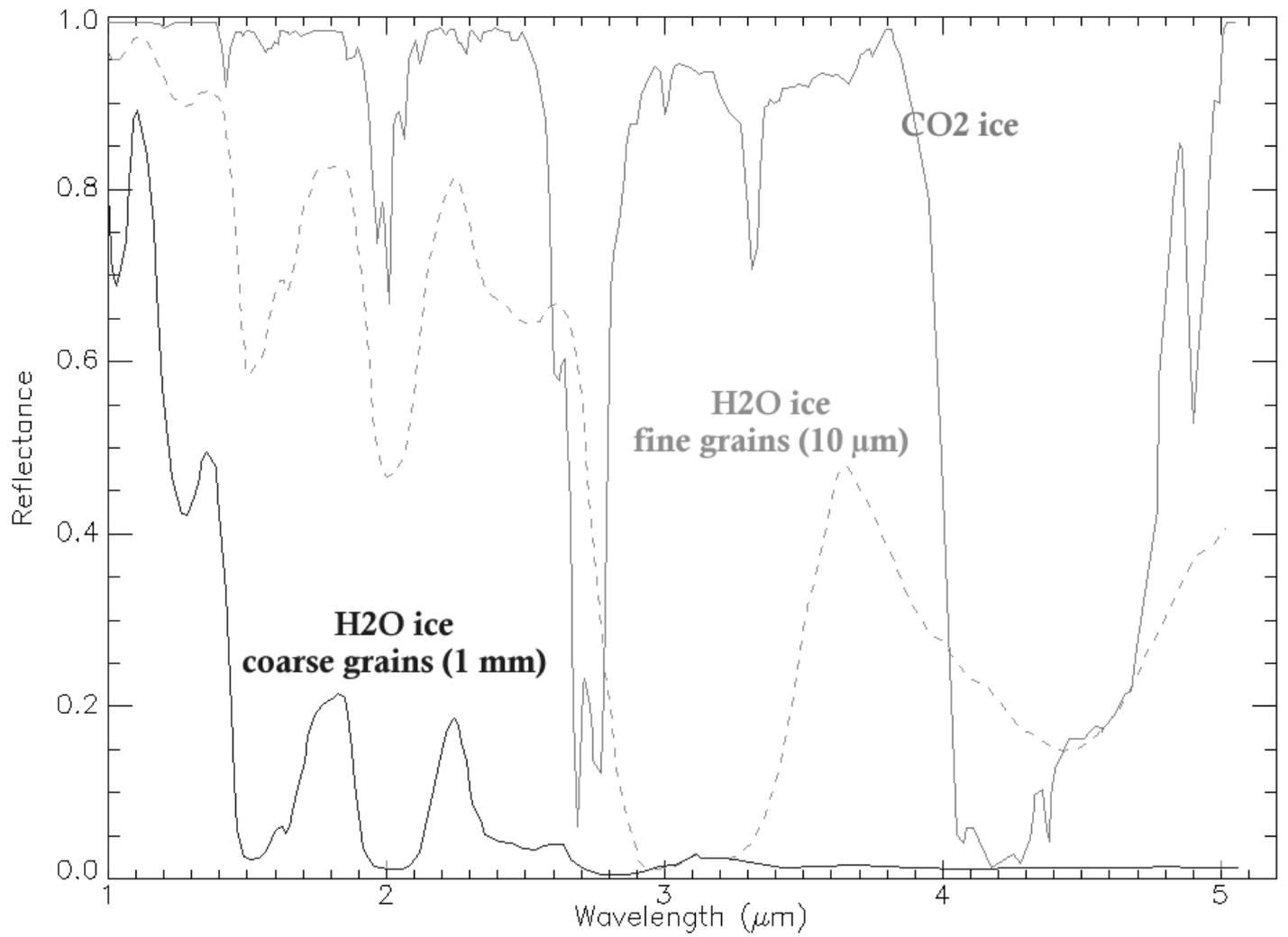

**Figure 11**

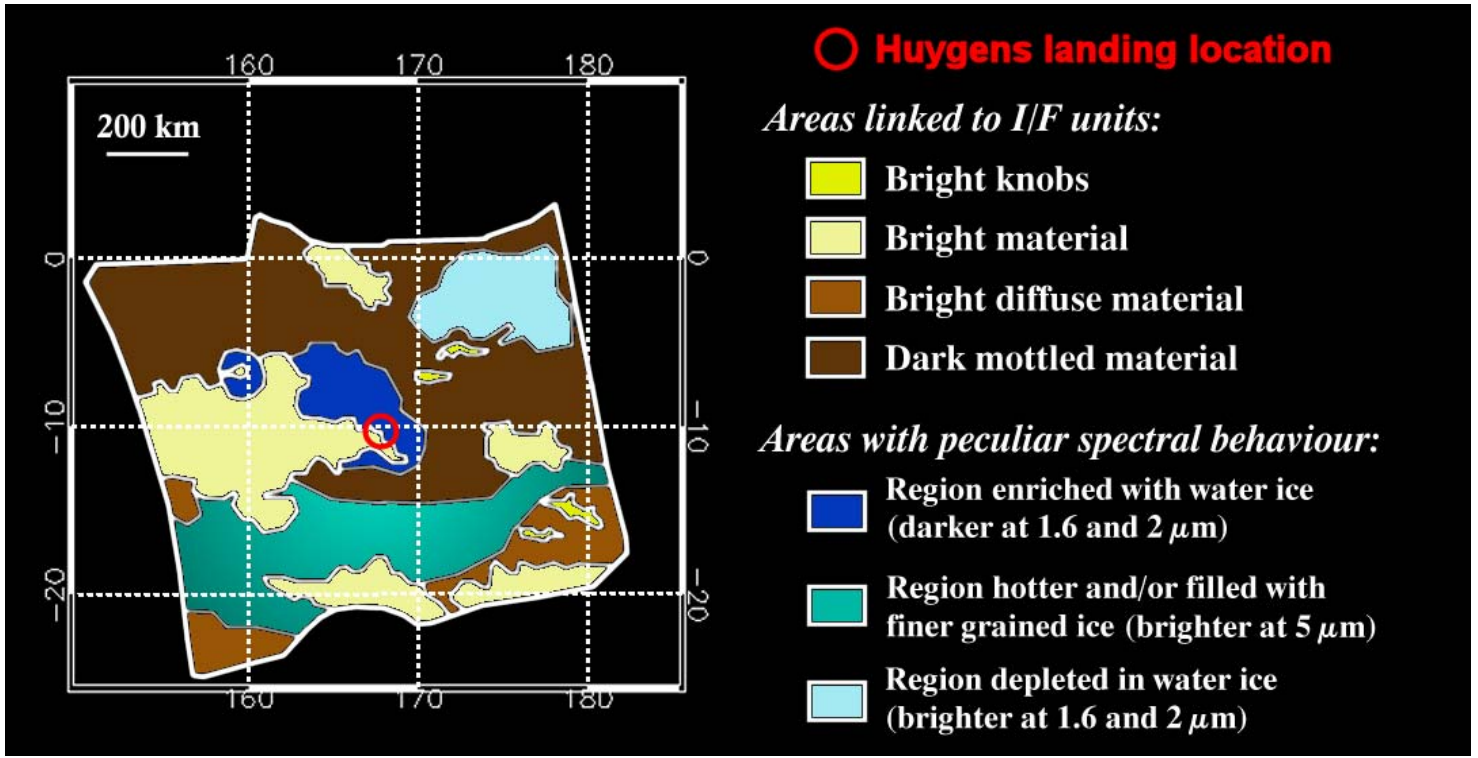

Figure 11